\documentclass[11pt]{article}
\usepackage{float}
\usepackage{array}
\usepackage{cool}
\usepackage{bbold}
\usepackage{eulervm}
\usepackage{hyperref}
\usepackage{tabularx}
\hypersetup{colorlinks=true,linkcolor=blue,citecolor=blue}
\let\oldmb\mathbold
\protected\def\mathbold{\oldmb}
\usepackage{amsmath}
\usepackage{todonotes}
\usepackage{graphicx}
\usepackage{subfig}
\usepackage{color}
 \usepackage{multicol}
\begin{document}
\title{\bf Plasmonics in the visible domain for a one-dimensional truncated photonic crystal terminated by graphene: sensing beyond Dirac point's approximation}

\author{A. Alidoust Ghatar, D. Jahani \footnote{\href{dariush110@gmail.com}{dariush110@gmail.com}}, W. Fritzche and F. Garwe}
\maketitle {\it \centerline{
\emph{Materials and Energy Research Center, Tehran, Iran}}}\maketitle {\it \centerline{
\emph{Leibniz-Institut for Photonische Technologien, Jena, Germany.}}}

\begin{abstract}
 \emph{Visible surface plasmon resonances (SPRs) could be excited by TE wave polarization in one-dimensional photonic crystals (PCs) coated by a graphene layer under the Kretschmann configuration. In this work, the plasmonic Bloch wave properties beyond Dirac points in a one-dimensional graphene-based photonic sensing structure have been numerically studied. We demonstrate that emergent plasmonic dips in the reflectance spectra of the suggested photonic device in the visible region exhibit tunable characteristics upon modulation of the chemical potential and the hopping parameter. The sensitivity of the sensor in the visible domain has been numerically evaluated and also was compared with those considering the surface plasmon resonances in the terahertz regime.}
 \end{abstract}
\vspace{0.5cm} {\it \emph{Keywords}}: \emph{Surface states, Graphene; Plasmonics; Visible domain; Photonic crystal}

\begin{multicols}{2}
\section{ Introduction}

\emph{Plasmonics is a rapidly growing field that has attracted intensive attention in physics, biology, chemistry, medical diagnostics, organic chemical detection, etc. due to highly extraordinary sensitive specifications of surface states [1,2]. Surface plasmons (SPs) are propagating modes which could be created when a polarized light strikes an electrically conducting surface at the interface between two media, a conductor (mostly metal) and a dielectric [3]. The necessary condition for occurring this resonance is matching the phase of the incoming light and the natural frequency of the material [4,5]. At this resonance condition, a sharp dip at a particular wavelength appears in the reflection spectrum and the intensity of the reflected light significantly shows to be reduced. Then, an unknown analyte could be detected by observing the variation of this dip in an optical sensing device [5,6,7]. We note that dip variations are extremely sensitive to changes in the refractive index of the dielectric medium and this property could be used as a sensing method especially as an optical biosensor [8,9]. The first sensing application of surface plasmonic resonance (SPRs) technique was reported by Liedberg et al. in 1983 [6]. Since then, SPRs have received much attention due to their extremely sensitive performance in optics. In order to evaluate the quality and the performance of an optical biosensor relevant parameters must be considered, for instance, the sensitivity, the depth and the width of the dip, size, the limit of detection and the cost [10,11,12]. Here, it should be noted that the earliest SPR sensing devices were bulky in size and mechanically unstable making them not suitable for the remote sensing applications [4,5,6,7]. To overcome mentioned limitations, manufacturing low-cost optical structures with high confinement characteristics for surface states propagation known as Photonic crystals (PCs) is of much interest [6,7,8,9,10,11,12,13].}

\emph{PCs are artificial optical structure with a periodic change in their refractive index which could create a forbidden frequency region due to various reflections and refractions that occur upon lights propagations [14]. Moreover, by considering plasmonic properties of surface states in a PC, designing optical structures in nanometer size at or less than the input wavelength will be possible. Therefore, scaling down to nanometer size for the PCs is possible by considering the sensing properties of the visible light.}

\emph{Furthermore, the choice of material for plasmonic waves have highly affected on sensing performance and better accuracy of SPR sensors. Therefore, a metallic component that carries large amounts of free electrons are required in these devices. In fact, gold and silver are mostly used metals in SPR sensing applications since gold is chemically stable in many environments and shows a large resonance peak shift. However, this peak resonance band becomes broader due to the large coefficient absorption which correspondingly reduces the accuracy of the analyte detection. On the contrary, silver with intrinsically higher sensitivities than gold shows lower loss and sharper resonance peaks compared to the other plasmonic materials. Unfortunately, silver shows rather quickly lost SPR properties with the contact of the ambient air because it is not chemically stable and could be oxidized easily [15,16,17,18]. Hence, it could be coated with an utra-thin graphene layer which could also absorb gas molecules on its surface [19,20]. Regardless of the use of noble metals for plasmonic materials, these conventional materials have their limitations. In fact, their optical properties are hardly tunable, due to large Ohmic losses cause poor propagating distances for plasmon oscillations. Therefore, we need a structure or materials that compensate for these limitations. Now, because of the outstanding properties of graphene which yields a negative imaginary part in optical conductivity in different frequency ranges from terahertz to visible light, it could be considered as an excellent alternative to the conventional plasmonic structures and noble metals [21,22,23,24,25].}

\emph{Graphene is a gapless flat monolayer of graphite with very high electron mobility whose transport characteristics and optical conductivity $(\sigma_{g})$ can be tuned by either electrostatic or magneto-static gating or chemical doping [26,??,27]. In addition, graphene has also been demonstrated to be a good candidate in optoelectronics with potential applications in photodetection, optics and plasmonics [28]. Moreover, advantages of graphene-based plasmonics are the presence of high carrier concentrations, ultra-relativistic speed for Dirac fermions, high confinement for surface states with longer propagation distance, flexible feature, adjustable electro-optical properties, nonlinear optics and large high-intensity electromagnetic field [29,30,31,32]. We should note that most of the interesting properties of graphene reflects back from its linear energy dispersion near two inequivalent points in the momentum space named as Dirac points. However, beyond Dirac points graphene's energy relation exhibits nonlinear properties which must be taken into account for obtaining its surface conductivity in the visible range [33]. Therefore, in the visible region, we need to consider optical excitations beyond Dirac points in order to investigate the surface states in graphene in the nonlinear regime.}

\emph{We note that in the previous works infrared light has been used while the wavelength of this region is relatively large, to be used for designing on-chip nano-scale optical devices [20,34,35]. In this paper, we investigate the sensing properties of SPs in the visible region allowing us to down the size of photonic crystal layers in a nanometer-scale that is comparable to the thickness of the graphene layer. To best of our knowledge, visible SPs corresponding to the optical excitations beyond Dirac points in graphene's spectrum have been reported up to this stage.}

\emph{This paper is organized into the following sections: The model is introduced in section 2. Following this section, we express the optical conductivity of the graphene sheet in the visible region, and by the use of the transfer matrix for the propagation of light among photonic crystals. Then numerical calculations are addressed in section 3. In the end, the conclusion is introduced in section 4.}

\section{Structure and Theoretical Modeling}
\emph{In order to compute the plasmon dispersion in our photonic structure, we first employ the transfer matrix method. It is assumed that light is polarized in the y-direction and propagating in the z-direction. The electromagnetic fields of the transverse electric (TE) polarized wave can be written
\begin{eqnarray}
 H_{1y}=(a_{1}e^{ik_{1z}z}+b_{1}e^{ik_{1z}z})e^{ik_{1x}x} \hspace{10mm} z<0,
  \end{eqnarray}
\begin{eqnarray}
 H_{2y}=(a_{2}e^{ik_{2z}z}+b_{2}e^{ik_{2z}z})e^{ik_{2x}x} \hspace{10mm} z>0.
  \end{eqnarray}
Here, $a_{i}$ and $b_{i} (i=1,2)$ are the field coefficients, $ k_{ix}(k_{iz}) $ is the $x(z)$ component of the wave-vector $k_{i}=\sqrt{\varepsilon_{i}}\omega/c $ in which $\omega$ represents the angular frequency and $c$ is the light speed in the vacuum. In the parenthesis on the right side, the first (second) term stands for propagation waves along z-direction.
We, then, relate the electric $E_{1}$ and $E_{2}$ and magnetic fields at the interface by using the following boundary conditions.
\begin{eqnarray}
 n\times(E_{2}-E_{1})|_{z=0}=0,
  \end{eqnarray}
  \begin{eqnarray}
 n\times(H_{2}-H_{1})|_{z=0}=J,
  \end{eqnarray}
 where $n$ is the unit of the surface normal and $J$ is the surface current density of the graphene. Subsequently, by applying the above boundary conditions at $z=0$ [36],
  \begin{eqnarray}
 \frac{k_{1z}}{\varepsilon_{1}}(a_{1}-b_{1})-\frac{k_{2z}}{\varepsilon_{2}}(a_{2}-b_{2})=0,
  \end{eqnarray}
  \begin{eqnarray}
 (a_{1}+b_{1})- (a_{2}+b_{2})=J_{x}.
  \end{eqnarray}
It should be pointed out that $J$ can be obtained from Ohm's law, that is,
\begin{eqnarray}
 J_{x}=\sigma E_{x}|_{z=0}=\frac{\sigma k_{2z}}{\varepsilon_{0}\varepsilon_{2}\omega}(a_{2}-b_{_{2}}),
  \end{eqnarray}
where $ \varepsilon_{0}$ is the vacuum permittivity and $\sigma$ is the optical conductivity of graphene in the visible region that follow below. It this important to mention that in the below equation the well-known Kubo formula is used, both inter-band and intra-band transitions are considered. The direction of the external field and optical conductivity is assumed the same [37].
\begin{equation}\begin{split}
\sigma(\omega)&=\\
&\frac{e^2 t^2 a^2\pi}{12\hbar^2\omega A_{c}}\frac{\hbar \omega}{t^2 \pi^2}\frac{1}{\sqrt{(1+\frac{\hbar\omega}{2t})^2-[(\frac{\hbar\omega}{2t})^2-1]^2/4}}\times\\&
\left[18-\frac{(\hbar\omega)^2}{t^2}\right]\left[\tanh(\frac{\beta(\hbar\omega+2\mu)}{4})+\tanh(\frac{\beta(\hbar\omega-2\mu)}{4}) \right]\\ &
K\left( \frac{\frac{4\hbar\omega}{2t}}{(1+\frac{\hbar\omega}{2t})^2-\frac{[(\frac{\hbar\omega}{2t})^2-1]^2}{4}}\right)+i Im(\sigma(\omega))
\end{split}.\end{equation}
with
\begin{equation}\begin{split}
Im(\sigma(\omega))&=\\
& \frac{e^2}{\hbar^2\pi\omega}\left( \mu-\frac{2}{9}\frac{\mu^3}{t^2}\right)-\frac{e^2}{4\pi\hbar}\log\frac{\vert\hbar\omega+2\mu\vert}{\vert\hbar\omega-2\mu\vert}\\&
-\frac{e^2 \hbar \omega^2}{144\hbar\pi t^2} \log\frac{\vert\hbar\omega+2\mu\vert}{\vert\hbar\omega-2\mu\vert}
\end{split}.\end{equation}
Here $\mu$, $\beta=1/(K_{B}T)$  and $t$  are the chemical potential, the Boltzmann constant and hopping parameter which is the energy for an electron to jump into its nearest neighbors, is in the order of $3 \ eV$, respectively. The lattice constant and the area of the unit cell are $a=1.42\ {\AA}$  and $A_{C}=(3\sqrt3 a^{2}/2)$ , respectively. The first kind of elliptic integral $K(r)$ is:
\begin{equation}
K(m)=\int_{0}^{1} dr \left[ \left( 1-r^{2}\right) \left( 1-m r^{2}\right)\right]^{-1}.
\end{equation}
with $m=\frac{\frac{4\hbar\omega}{2t}}{(1+\frac{\hbar\omega}{2t})^2-\frac{[(\frac{\hbar\omega}{2t})^2-1]^2}{4}}$ .
The coefficients $a_{1}$ and $b_{1}$ can be related to $a_{2}$ and $b_{2}$ by the $2\times 2$ transmission matrix $D_{1\rightarrow2}$ combining equations (1-5)-(1-6) as follows:
\begin{equation}
 \binom{a_{1}}{b_{1}}=D_{1\rightarrow2}\binom{a_{2}}{b_{2}},
  \end{equation}
 where
 \begin{equation}
 D_{1\rightarrow2}=\frac{1}{2}
 \begin{pmatrix}
  1+\eta_{s}+\xi_{s} & 1-\eta_{s}+\xi_{s} \\
  1-\eta_{s}-\xi_{s} & 1+\eta_{s}-\xi_{s}
\end{pmatrix},
  \end{equation}
with the parameters $\eta_{s}$ and $\xi_{s}$ provided by:
\begin{equation}
 \eta_{s}=\frac{k_{2z}}{k_{1z}}, \hspace{10mm} \xi_{s}=\frac{\sigma \mu_{0}\omega}{k_{1z}}.
  \end{equation}
   Now, we can examine light propagation in a homogeneous medium. The electric or magnetic field at $z+\Delta z$ can be related to those at the $z$ position by a $2\times 2$ propagation matrix
\begin{equation}
 P(\Delta z)=
 \begin{pmatrix}
  e^{-ik_{z}\Delta z} & 0 \\
  0 & e^{ik_{z}\Delta z}
\end{pmatrix}.
\end{equation}
For a stack of $N$ graphene layers, the transfer matrix can be obtained by transmission matrices across different interfaces and propagation matrices in different homogeneous dielectric media. The two sets of field coefficients are then related by a  $2\times 2$ transfer matrix $M$, namely
\begin{equation}
 \binom{a_{1}}{b_{1}}=M\binom{a_{N+1}}{b_{N+1}}.
  \end{equation}
  with
  \begin{equation}
M=D_{1\rightarrow 2}P(d_{1,2})D_{2\rightarrow 3}P(d_{2,3})....P(d_{N-1,N})D_{N\rightarrow N+1}.
  \end{equation}
 Finally, the transmittance $(T)$ and reflectance $(R)$ can be calculated by using the matrix elements $M_{11}$  and $M_{12} $  as follows :
\begin{equation}
R=\Big|\frac{M_{21}}{M_{11}}\Big|^2 ,   \hspace{10mm}   T=\Big|\frac{1}{M_{11}}\Big|^2.                                                                                                                                 \end{equation}
 Now, using the above expression we are ready to calculate the transmission spectrum of our 1D photonic crystal. In the next section, therefore, the numerical results are expressed in more detail.}
\section{Result and discussion}
\emph{For our numerical approach, in this section, based on the transmission matrix method, a trunked photonic crystal coated with single-layer graphene with Kretschmann configuration [?] is assumed which is demonstrated schematically in Figure 1. It consists of two dielectric layers with constants $\varepsilon_{A}$ and $\varepsilon_{B}$ and thicknesses of $ d_{A}$ and $d_{B}$ arranged in the periodic pattern, respectively. As it is seen, the graphene layer is attached to the photonic crystal contact with a sample whose refractive index changes, and a prism with refractive index $n_{p}=1.77$   at the entrance side of the proposed sensing device. The surface plasmonic wave excitations can be observed by a dip appearing in the reflection spectra of the photonic structure. We choose $Si$ for the $A$, $SiO_{2}$ for the material $B$  and the period of the crystal is $N = 9$. The region of the transmission spectra which is considered to be in the visible zone $(1-2 \ eV)$ requires choosing proper geometrical parameters for thicknesses of the layers. Therefore, the thickness of A and B layers are chosen to be $d_{A}=d_{B}= 200 \ nm$. We note that the temperature in all of our calculations is assumed to be $T=300\ K$ which is the well-known room temperature.}

\emph{Also, the dielectric constants of layers are $12.74$ and $2.16$ for $Si$ and $SiO2$, respectively. Nore that, the imaginary parts of the refractive indices of dielectric layers are considered to be zero in this work. Now, for exciting the surface states we consider visible light to be incident at greater angles than $\theta_{c}=sin^{-1}(n_{s}/n_{p} )$ for which reflection is perfect. Therefore, one can detect plasmonic surface states have appeared as a dip in this reflection spectrum near the band-gap-edge in our numerical calculations.}

\emph{As our main aim in this section, we will investigate the dip properties in the reflection spectra of the introduced truncated photonic crystal by using (2-16). At first, we present the reflection spectra of the proposed structure without a graphene layer based on the parameters given above, as shown in Fig. 2. A small dip (blue solid line) appears in the spectrum at $\lambda=637.6 \ nm$ wavelength and $\theta=34.4$ degree. Interestingly, by introducing a graphene layer into the supposed structure with the relatively low chemical potential of about $\mu=0.1 \ eV$ and the hopping parameter $t=3.1 \ eV$ no dip is appeared in the dispersion spectrum (Green solid line). The situation changes if one goes beyond Dirac points by considering larger values for the chemical potential for graphene so that a deep dip (dotted red line) in reflectance spectra of the proposed PC appears in $\lambda=635.6 \ nm$ , $\mu=1.145 \ eV$ and $t=3.1 \ eV$ as shown in Fig. 2.}

\emph{To proceed, as a matter of more comparison, in Fig. 3 we show both the reflection and transmission profiles of the proposed sensing structure terminated by single-layer graphene. As it is evident from this figure the light is completely absorbed and we have no transmitted light in the corresponding spectrum which indicates well the emergence of a plasmonic surface state at the interface between graphene and the sample.}

\emph{Moreover, in Fig. 4, the obtained resonant dip is illustrated for two situations, one for the modulation of the incident angles in constant wavelength $\lambda=635/ nm$ and the other(inset figure) for a specific interval for the wavelength in constant incident angle $\theta=34.44$  degree which the plasmonic wave is centered around $635 \ nm$. It is clear that one gets a more narrow dip which indicates proper physical parameters selectivity and suitable for serving as a gas sensor in the visible region. Then, to evaluate the angular and wavelength sensitivity (S) of the sensor, we need to have the slope of these diagrams $S_{\theta}=\Delta\theta/\Delta n$  and $S_{\lambda}=\Delta\lambda/\Delta n \ (nm)$  which indicate the angular shift and wavelength peak shift of surface plasmons relative to the refractive index changes, respectively. In Fig. 5, the reflectance of the optical sensing layered device is plotted as a function of the wavelength and also the angle of incidence. Then the angular sensitivity of the sensor is, therefore, computed at $\lambda=635.6 \ nm$ with the refractive indices ranging from 1 to 0.01 in an interval of 0.002 in fig 5(.a) and (c). Also, the wavelength sensitivity is calculated at $\theta=34.44$ degree with refractive indices ranging from 1 to 0.001 in an interval of 0.0002 in figure 5(b) and (d). Moreover, it is evident that the variations of the resonant angle due to variations of refractive indices show a linear behavior and, therefore, the sensitivity is equal to $S_{\theta}=35.53 \ \circ/RIU$ (Refractive Units Index). Similarly, the spectra in Fig. 5(b) and (d) demonstrate that the dip wavelength sensitivity is $S_{\lambda}=1402 \ nm/RIU$ and the variation of the resonance wavelength is almost linear and red-shifted which indicates both sensitivities are higher than the results reported previously [38,39].}

\emph{In the following, the effect of increasing the chemical potential and the hopping energy on the refractive spectra and the sensitivity of the device is investigated. Hence, we investigate the dip modulation by considering the hopping energies of graphene, $t = 2.5 \ eV$ and $t = 3.1 \ eV$ for different refractive values. As is shown in Fig. 6, by increasing the hopping parameters which means that the gas molecules in the air are absorbed on graphene's surface shifts the refractive spectra without affecting the angular and wavelength sensitivities meaning that $S_{\theta}=35.42\  \circ/RIU$ and $S_{\lambda}=1399\  nm/RIU $ remain almost constant. Therefore, our sensor is sensitive to the different kinds of gas molecules absorption on graphene's surface without any change in sensing performance which is of most interest in sensing applications.}

\emph{In the end, since the chemical potential, $\mu$, significantly could affect the charge carrier's concentration of graphene, its influence on SPRs can also be analyzed. In Fig. 7 (a),(b) the refractive spectra is shown to be red-shifted without affecting much the sensitivities as also reported for the infrared region in [34].}

\section{Conclusion}

 \emph{In this work, we addressed optical sensing properties of SPRs for a 1D truncated PC coated by graphene in the visible region of the light's frequency. In particular, we have numerically calculated the spectra for the suggested structure in order to analyze the sensing properties of an emergent dip in reflectance due to the absorption of SPRs in graphene beyond its usual linear dispersion. As it was discussed these relatively high domains of the frequency for photons allows the fabricating of nanoscale layered structure which could be employed in on-chip sensing technology. Then, the refractive spectra and sensitivity of the proposed sensing device has been presented by the transfer matrix method.}
 \emph{The most significant result of this work is the emergence of highly narrow surface plasmons in the visible regions of the spectrum for graphene beyond its usual Dirac points approximation. Moreover, the effect of changing the refractive index of the analyte on the plasmonic dip in the reflectance spectra of the optical sensor was investigated. Our results indicated that $S_{\theta}=35.53\  \circ/RIU$ and $S_{\lambda}=1402\  nm/RIU $ for the angular and wavelength sensitivity, respectively meaning the low-loss and narrow width dip comparable to the previous works [38,39]. The role of hopping energy and the chemical potential also analyzed which shows that the plasmonic spectra are shifted SPRs propagations. In fact, with regard to the prospect of graphene, PCs and plasmonic in the future, the proposed SPR sensor may be considered an ideal candidate for sensing purposes in the visible domain.}

\end{multicols}
\begin{figure}
\begin{center}
\includegraphics[width=18cm]{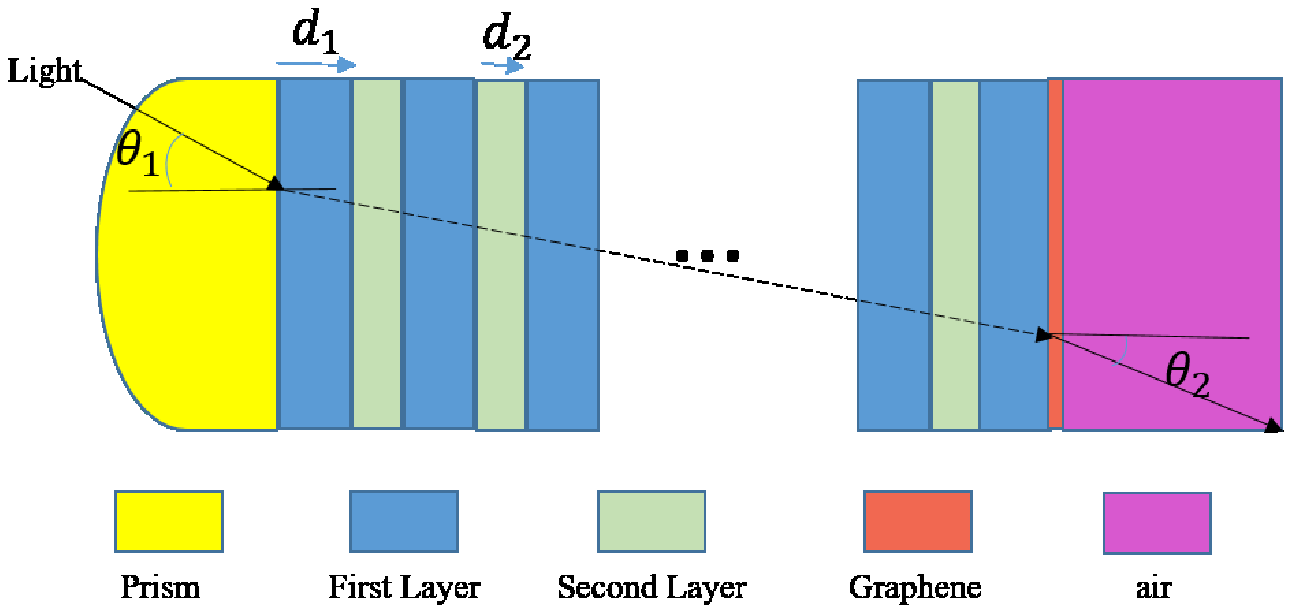}
\caption{ Schematic diagram of the designed structure in the presence of a prism with $n=1.77$ and the period $N=9$ coated by a single-layer graphene.}
\end{center}
\end{figure}
 \begin{figure}
\begin{center}
\includegraphics[width=18cm]{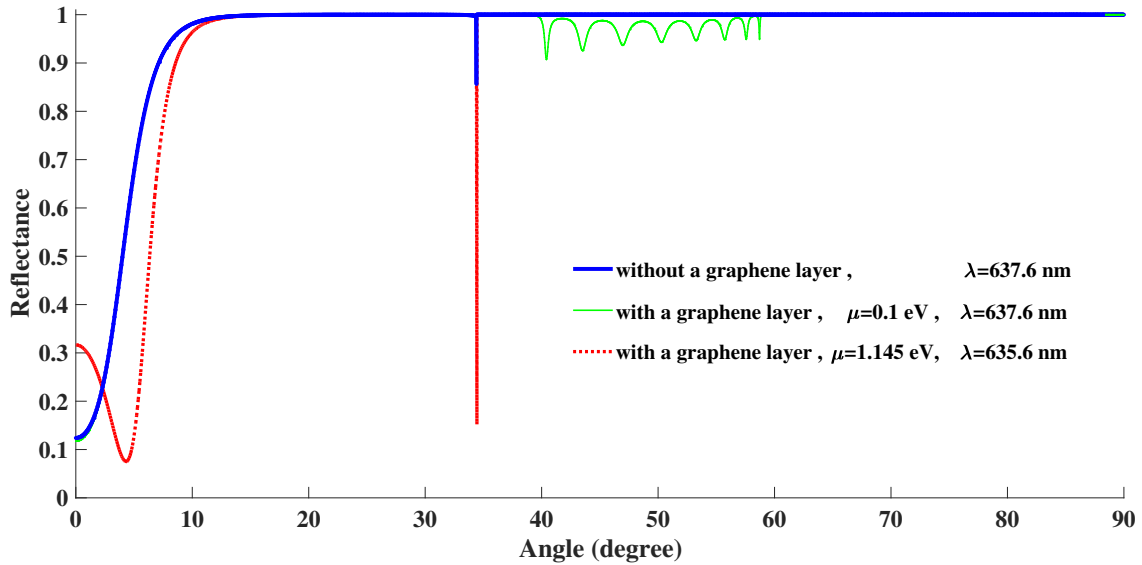}
\caption{ Reflection spectrum for the SPRs in the visible domain of the suggested sensing structure as a function of the incident angle. The blue solid line represents the graphene-less structure for $\lambda=637.6 \ nm$ and $T=300\ K$. The green line represents the structure with a graphene layer for $\lambda=637.6 \ nm$ , $\mu=0.1 \ eV$ and $t=3.1 \ eV$. The red dotted line is for the graphene-based structure for $\lambda=635.6 \ nm$ $\mu=1.145 \ eV$ and $t=3.1 \ eV$.}
\end{center}
\end{figure}
 \begin{figure}
\begin{center}
\includegraphics[width=18cm]{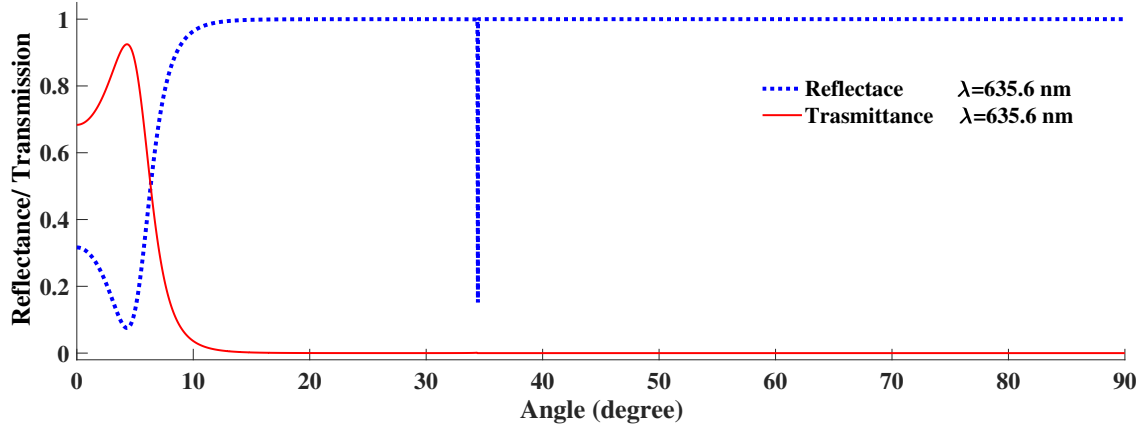}
\caption{ Reflection and the transmission spectra for detecting SPR dip in the case of $\mu= 1.145 \ eV$ and $t =3.2 \ eV$ for $\lambda=635.6 \ nm$. As it is clear, no transmitted light is obverted in the transmission spectra of the SPR photonic sensing structure while a plasmonic dip is emerged at the wavelength $\lambda=635.6 \ nm$.}
\end{center}
\end{figure}
 \begin{figure}
\begin{center}
\includegraphics[width=18cm]{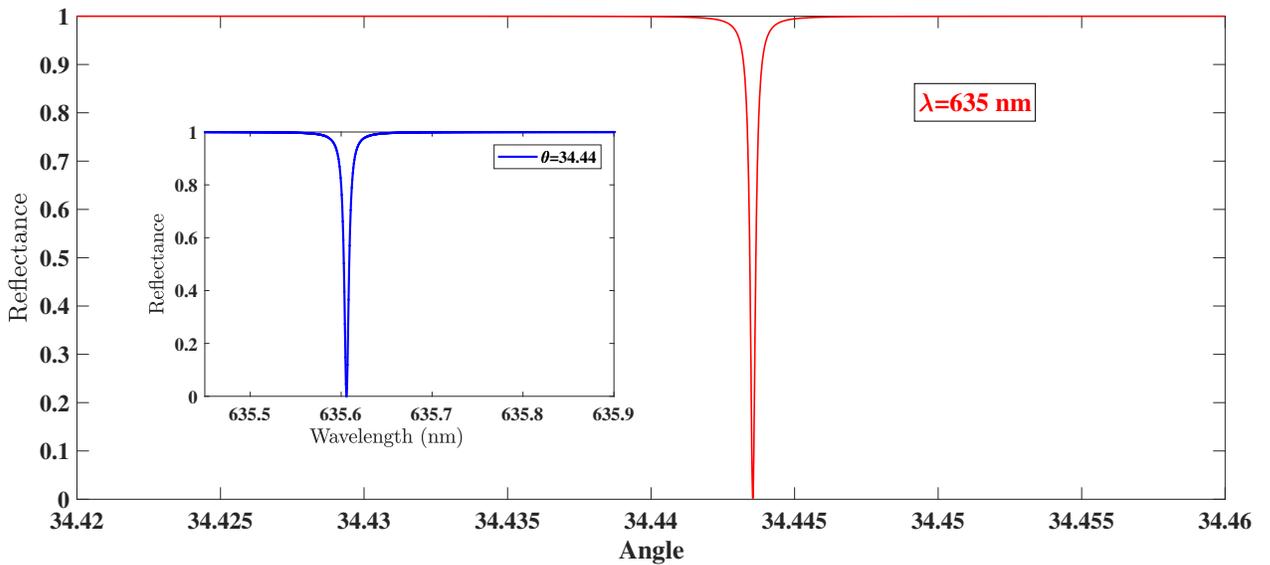}
\caption{ Results of numerical calculations for detecting a Plasmonic dip in the reflectance spectrum of the truncated PC as a function of the incident angles at constant wavelength around $635 \ nm$, and the inset showes the reflectance spectrum as a function of the wavelength centered around $635 \ nm$ in constant incident angle $\theta=34.44$ degree.}
\end{center}
\end{figure}
 \begin{figure}
\begin{center}
\includegraphics[width=18cm]{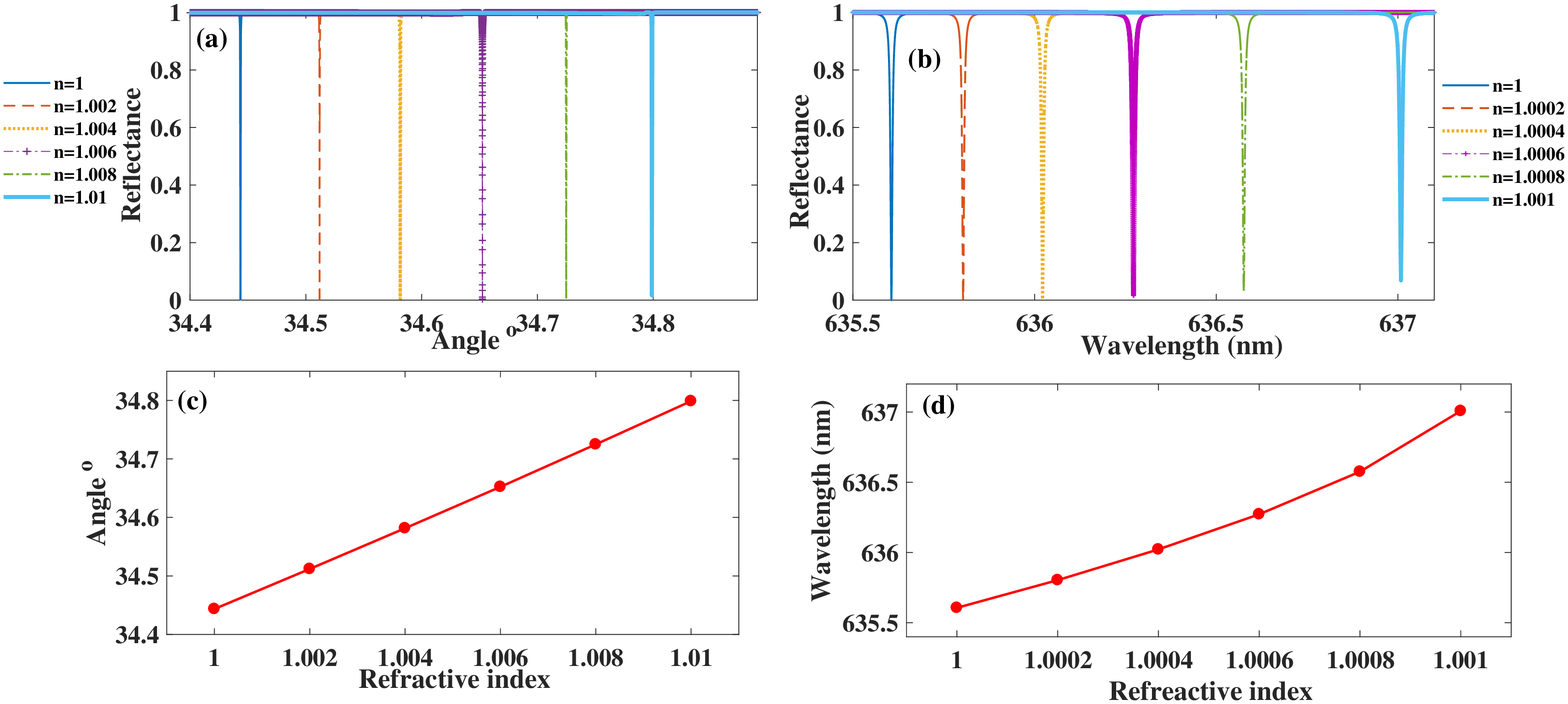}
\caption{ (a) and (c): the angle variations of reflectance as a function of the wavelength under for different refractive indices with $\mu=1.145 \ eV$ and $t=3.1 \ eV$. (b) and (d): Wavelength's variation for the reflectance as a function of the incident angle for different values for refractive indices for $\theta=34.44 ^{o}$ and $t=3.1 \ eV$ keeping all the parameters the same as that in Fig. 2.}
\end{center}
\end{figure}
 \begin{figure}
\begin{center}
\includegraphics[width=18cm]{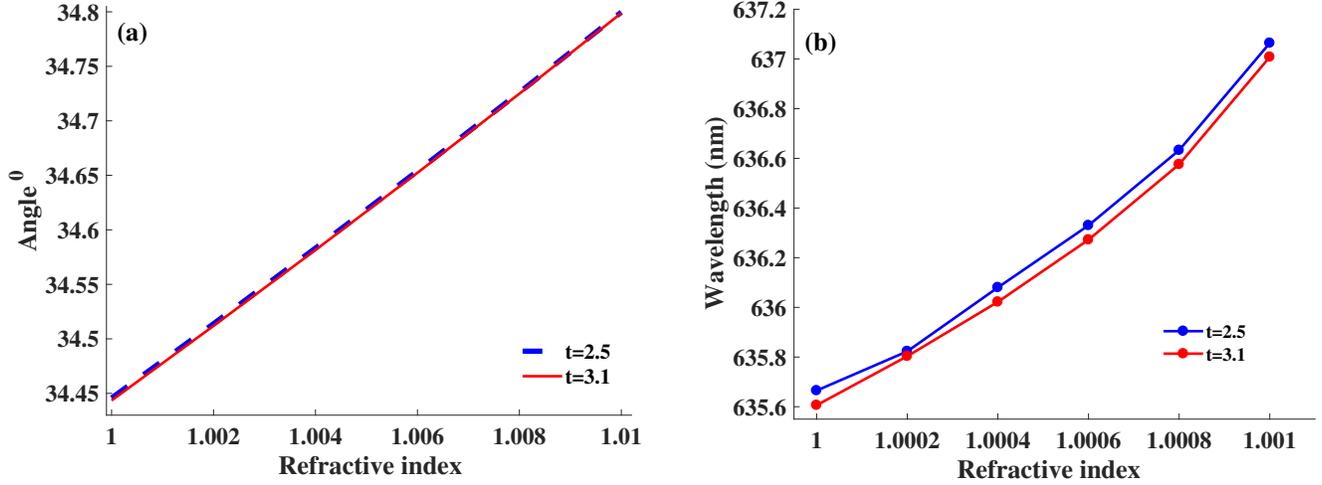}
\caption{ (a) Variation of the plasmonic dip for surface states as a function of the incident angle and refractive indices for $\lambda=635.6 nm$ and, (b) Variation of the plasmonic dip as a function of the wavelength for the different values of refractive indices with $\theta=34.44 ^{o}$, the hopping energy $t=2.5 \ eV$ and $3.1 \ eV$. Here the chemical potential and wavelength are set to be $\mu=1.145$ and $\lambda = 635.6 nm$, respectively.}
\end{center}
\end{figure}
 \begin{figure}
\begin{center}
\includegraphics[width=18cm]{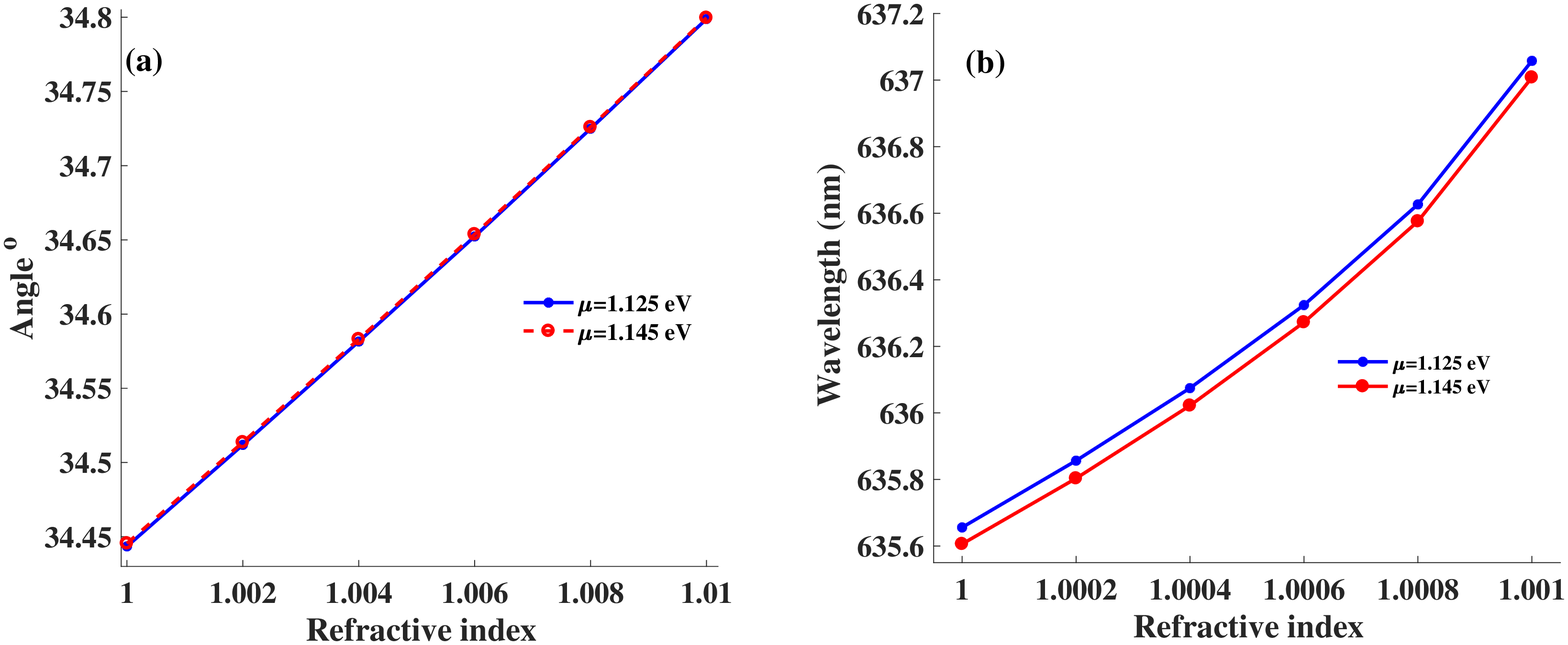}
\caption{ (a) Variation of the plasmonic dip for the incident angle as a function of the refractive indices for $\lambda=635.6 \ nm$ and, (b) wavelength's variation of the SPRs as a function of the refractive index for $\theta=34.44 ^{o}$, $\mu = 1.125$ and $1.145 \ eV$ and a constant constant hopping parameter $t = 3.1 \ eV$.}
\end{center}
\end{figure}


\begin{thebibliography}{99}
\bibitem{1} Y. B. Zheng, B. Kiraly, P. S.Weiss, and T. J.  Huang, (2012). Molecular plasmonics for biology and nanomedicine. Nanomedicine, 7(5), 751-770.
\bibitem{2} Y. Saylan, S. Akgonullu, and A. Denizli, (2020). Plasmonic Sensors for Monitoring Biological and Chemical Threat Agents. Biosensors, 10(10), 142.
\bibitem{3} I. R. Hooper, and J. R. Sambles, (2002). Dispersion of surface plasmon polaritons on short-pitch metal gratings. Physical Review B, 65(16), 165432.
\bibitem{4} B. D. Gupta, and R. K. Verma, (2009). Surface plasmon resonance-based fiber optic sensors: principle, probe designs, and some applications. Journal of sensors.
\bibitem{5} J. Homola, S. S. Yee, and G. Gauglitz, (1999). Surface plasmon resonance sensors. Sensors and actuators B: Chemical, 54(1-2), 3-15.
\bibitem{6} B. Liedberg, C. Nylander, and I. Lunstrom, (1983). Surface plasmon resonance for gas detection and biosensing. Sensors and actuators, 4, 299-304.
\bibitem{7} X. Yang, Y. Lu, B. Liu, and J. Yao, (2017). Analysis of graphene-based photonic crystal fiber sensor using birefringence and surface plasmon resonance. Plasmonics, 12(2), 489-496.
\bibitem{8} R. S. Zheng, Y. H. Lu, Z. G. Xie, J. Tao, K. Q. Lin, and H. Ming, (2008, November). Surface plasmon resonance sensors based on polymer optical fiber. In 2008 1st Asia-Pacific Optical Fiber Sensors Conference (pp. 1-4). IEEE.
\bibitem{9} S. biosensors for use in the diagnostics of malignant and infectious diseases. Laser Physics Letters, 15(6), 065602.Firdous, S. Anwar, and R. Rafya, (2018). Development of surface plasmon resonance (SPR).
\bibitem{10} K. V. Sreekanth, S. Zeng, K. T. Yong, and T. Yu, (2013). Sensitivity enhanced biosensor using graphene-based one-dimensional photonic crystal. Sensors and Actuators B: Chemical, 182, 424-428.
\bibitem{11} X. Fan, I. M. White, S. I. Shopova, H. Zhu, J. D. Suter, and Y. Sun, (2008). Sensitive optical biosensors for unlabeled targets: A review. analytica chimica acta, 620(1-2), 8-26.
\bibitem{12} A. Panda, P. D. Pukhrambam, and G. Keiser, (2020). Performance analysis of graphene-based surface plasmon resonance biosensor for blood glucose and gas detection. Applied Physics A, 126(3), 1-12.
\bibitem{13} M. Hasan, S. Akter, A. A. Rifat, S. Rana, and S. Ali, (2017, March). A highly sensitive gold-coated photonic crystal fiber biosensor based on surface plasmon resonance. In Photonics (Vol. 4, No. 1, p. 18). Multidisciplinary Digital Publishing Institute.
\bibitem{14} E. Yablonovitch, (1987). Inhibited spontaneous emission in solid-state physics and electronics. Physical review letters, 58(20), 2059.
\bibitem{15} M. Kanso, S. Cuenot, and G. Louarn, (2007). Roughness effect on the SPR measurements for an optical fibre configuration: Experimental and numerical approaches. Journal of Optics A: Pure and Applied Optics, 9(7), 586.
\bibitem{16} J. N. Dash, and R. Jha, (2014). SPR biosensor based on polymer PCF coated with conducting metal oxide. IEEE Photonics Technology Letters, 26(6), 595-598.
\bibitem{17} E. K. Akowuah, T. Gorman, H. Ademgil, S. Haxha, G. K. Robinson, and J. V. Oliver, (2012). Numerical analysis of a photonic crystal fiber for biosensing applications. IEEE Journal of Quantum Electronics, 48(11), 1403-1410.
\bibitem{18} Y. Lu, C. J. Hao, B. Q. Wu, M. Musideke, L.  C. Duan, W. Q. Wen, and J. Q. Yao, (2013). Surface plasmon resonance sensor based on polymer photonic crystal fibers with metal nanolayers. Sensors, 13(1), 956-965.
\bibitem{19} W. Qin, S. Li, Y. Yao, X. Xin, and J. Xue, (2014). Analyte-filled core self-calibration microstructured optical fiber based plasmonic sensor for detecting high refractive index aqueous analyte. Optics and Lasers in Engineering, 58, 1-8.
\bibitem{20} L. Wu, H. S. Chu, W. S. Koh, and E. P. Li, (2010). Highly sensitive graphene biosensors based on surface plasmon resonance. Optics express, 18(14), 14395-14400.
\bibitem{21} A. N. Grigorenko, M. Polini, and K. S. Novoselov, (2012). Graphene plasmonics. Nature photonics, 6(11), 749-758.
\bibitem{22} F. H. L. MinovKoppensich, D. E. Chang, S. Thongrattanasiri, and F. G. de Abajo, (2011). Graphene plasmonics: A platform for strong light-matter interactions. Optics and Photonics News, 22(12), 36-36.
\bibitem{23} M. N. Gjerding, M. Pandey, and K. S. Thygesen, (2017). Band structure engineered layered metals for low-loss plasmonics. Nature communications, 8(1), 1-8.
\bibitem{24} Q. Bao, and K. P. Loh, (2012). Graphene photonics, plasmonics, and broadband optoelectronic devices. ACS nano, 6(5), 3677-3694.
\bibitem{25} P. R. West, S. Ishii, G. V. Naik, N. K. Emani, V. M. Shalaev, and , A. Boltasseva (2010). Searching for better plasmonic materials. Laser and Photonics Reviews, 4(6), 795-808.
\bibitem{26} S. Zhang, Z. Li, and F. Xing, (2020). Review of polarization optical devices based on graphene materials. International journal of molecular sciences, 21(5), 1608.
\bibitem{??} O. L. Berman, V. S. Boyko, R. Y. Kezerashvili, A. A. Kolesnikov, and Y. E. Lozovik, (2010). Graphene-based photonic crystal. Physics Letters A, 374(47), 4784-4786.
\bibitem{27} O. L. Berman, and R. Y. Kezerashvili, (2011). Graphene-based one-dimensional photonic crystal. Journal of Physics: Condensed Matter, 24(1), 015305.
\bibitem{28} P. Avouris, and M. Freitag, (2013). Graphene photonics, plasmonics, and optoelectronics. IEEE Journal of selected topics in quantum electronics, 20(1), 72-83.
\bibitem{29} T. Low, and P. Avouris, (2014). Graphene plasmonics for terahertz to mid-infrared applications. ACS nano, 8(2), 1086-1101.
\bibitem{30} Z. Zhen, and H. Zhu, (2018). Structure and properties of graphene. In Graphene (pp. 1-12). Academic Press.
\bibitem{31} Z. Xu, (2018). Fundamental properties of graphene. In Graphene (pp. 73-102). Academic Press.
\bibitem{32} X. Luo, T. Qiu, W. Lu, and Z. Ni, (2013). Plasmons in graphene: recent progress and applications. Materials Science and Engineering: R: Reports, 74(11), 351-376.
\bibitem{33} A. C. Neto, F. Guinea, N. M. Peres, K. S. Novoselov, and A. K. Geim, (2009). The electronic properties of graphene. Reviews of modern physics, 81(1), 109.
\bibitem{34} A. K. Sharma, and B. Kaur, (2018). Analyzing the effect of graphene's chemical potential on the performance of a plasmonic sensor in infrared. Solid State Communications, 275, 58-62.
\bibitem{35} J. Tang, Y. Ye, J. Xu, Z. Zheng, X. Jin, L. Jiang, ... and Y. Xiang, (2020). High-sensitivity terahertz refractive index sensor in a multilayered structure with graphene. Nanomaterials, 10(3), 500.
\bibitem{36} T. Zhan, X. Shi, Y. Dai, X. Liu, and J. Zi, (2013). Transfer matrix method for optics in graphene layers. Journal of Physics: Condensed Matter, 25(21), 215301.
\bibitem{37} T. Stauber, N. M. R. Peres, and A. K. Geim, (2008). Optical conductivity of graphene in the visible region of the spectrum. Physical Review B, 78(8), 085432.
\bibitem{?} E. Kretschmann, and H. Raether, (1968). Radiative decay of non-radiative surface plasmons excited by light. Z. Naturforsch. a, 23(12), 2135-2136.
\bibitem{38} A. V. Baryshev, and A. M. Merzlikin, (2014). Plasmonic photonic-crystal slabs: visualization of the Bloch surface wave resonance for an ultrasensitive, robust and reusable optical biosensor. Crystals, 4(4), 498-508.
\bibitem{39} X. J. Zou, G. G. Zheng, and Y. Y. Chen, (2018). Confinement of Bloch surface waves in a graphene-based one-dimensional photonic crystal and sensing applications. Chinese Physics B, 27(5), 054102.


\end{thebibliography}
\end{document}